\documentclass[]{article}

\usepackage[utf8]{inputenc}
\usepackage[english]{babel}
\usepackage{amsmath}
\usepackage{amsfonts}
\usepackage{amssymb}
\usepackage{dsfont}

\providecommand{\keywords}[1]
{
  \small	
  \textbf{\textit{Keywords---}} #1
}
\providecommand{\PACS}[1]
{
  \small	
  \textbf{\textit{PACS:}} #1
}

\begin{document}
\title{Categorical generalization of 
MacDowell-Mansouri gravity coupled to Kalb–Ramond fields
}
\author{Miguel A. Oliveira\footnote{Electronic address: masm.oliveira@gmail.com}
\\
             Grupo de Física Matemática,\\
              Faculdade de Ciências da Universidade de Lisboa,\\
			Campo Grande, Edifício C6, 1749-016 Lisboa, Portugal \\%
}
\maketitle

\begin{abstract}
In this work we generalize the MacDowell-Mansouri theory of gravity using strict 2-groups. To achieve this, we construct the categorical generalization of the $ISO(4,1)$ group, which we call the \emph{de Sitter 2-group}. We then proceed to generalize the MacDowell-Mansouri theory in two different ways. First, as a Yang-Mills-type theory, where the symmetry is  explicitly broken to obtain an Einstein-Cartan theory coupled to Kalb-Ramond fields. And second, by using the categorical generalization of the $BF$ theory, called 2-$BF$ theory, which after the addition of some symmetry breaking and constraint terms gives the same Einstein-Cartan theory coupled to Kalb-Ramond fields plus some extra topological terms.\vspace*{10pt}\\
\\
\end{abstract}

\section{Introduction}\label{intro}

The quantization of gravity is one of the most important problems in modern physics. To help solve this problem, an effort has been made to cast General Relativity (GR) in a form similar to the Yang-Mills theory (YM). The Ashtekar formulation of GR for example, taught us that we must view GR as the theory of connections instead of a theory of metrics \cite{MOphd}, thus bringing it closer to YM.

However a key difference between these two theories is the fact that in YM spacetime plays the role of a fixed background, where the dynamics takes place, whereas GR is a theory describing the dynamics of spacetime itself. 

The cosmological constant $\Lambda$, introduced by Einstein to create a static cosmological model, but most notably used today to account for the current accelerated expansion of the universe, is a key feature of GR and its origin is another important problem in modern physics. It is therefore important, to study  models of gravity with $\Lambda$, with a view to their subsequently quantization. 

One way to achieve this, is to use a gauge group containing the Lorentz group. MacDowell and Mansouri (MM) (see  \cite{MM,MMerr} for the original work, for recent applications see for example \cite{Lopez-Dominguez:2014yha,Berra-Montiel:2017led}, \cite{Alvarez:2021qbu} in supergravity, for a review and geometrical setting see \cite{Wise:2006sm}) proposed a theory  with a gauge group $G\supset SO(3,1)$. This group will depend on the signature of the metric and the sign of  $\Lambda$. For Lorentzian spacetime in four dimensions we have: the De Sitter  group $G=SO(4,1)$ for $\Lambda>0$ or the $G=SO(3,2)$ for $\Lambda<0$ \footnote{Whereas for Euclidean spacetime in four dimensions the group is: $G=SO(5)$ for $\Lambda>0$ or $G=SO(4,1)$ for $\Lambda<0$.}. Given the observational evidence we are interested in positive $\Lambda$, and in order to recover GR we use a Lorentzian spacetime manifold $M$ with signature  $(-+++)$, and therefore $G=SO(4,1)$ and the corresponding $SO(4,1)$ principal bundle $P$ over $M$.

In MM theory, the De Sitter Lie algebra $\mathfrak{so}(4,1)$ seen as a vector space, is broken in the following way:
\begin{equation}\label{Liesplitt}
\mathfrak{so}(4,1)\cong \mathfrak{so}(3,1) \oplus \mathds{R}^{3,1}\,.
\end{equation}

In components we have, the $\mathfrak{so}(4,1)$ connection $A^{IJ}$ where $IJ$ are De Sitter Lie algebra indices $I,J=0\ldots 4$, will split into the $\mathfrak{so}(3,1)$ (Lorentz) connection $A^{ab}=\omega^{ab}$ with $a,b=0\ldots 3$ and the co-tetrad $A^{a4}=\frac{1}{\ell} e^a$, 
\begin{equation}\label{MM_A}
	A^{IJ}{}_\mu=\left[\begin{array}{ccc}
	\omega^{ab}{}_\mu & & \frac{1}{\ell}e^a{}_\mu \\ 
	\\
	-\frac{1}{\ell}e^b{}_\mu & & 0
	\end{array}\right] \,,
\end{equation}
where, $\mu=0\ldots 3$ is a spacetime index\footnote{These index conventions will  be used throughout this work.} and $\ell$ is a parameter with dimensions of length.

The curvature of $A^{IJ}$ is:
\begin{equation}\label{FA}
	F^{IJ}=dA^{IJ}+A^I{}_K\wedge A^{KJ}\,,
\end{equation}
and it splits like:
\begin{eqnarray} 
	\label{Fab} F^{ab}&=&d\omega^{ab}+\omega^a{}_c\wedge\omega^{cb}-\frac{1}{\ell^2}e^a\wedge e^b\nonumber\\
		  &=&R^{ab}-\frac{1}{\ell^2}e^a\wedge e^b\,,\\
	\label{Fa4} F^{a4}&=& \frac{1}{\ell}\left( d e^a + \omega^a{}_c\wedge e^c\right)\nonumber\\
	      &=& \frac{1}{\ell}T^a\,,
\end{eqnarray}
where $R$ is the curvature of $\omega$ and $T$ is the torsion. 

The MM theory was originally formulated as a YM theory. However, simply using the curvature (\ref{FA}) in the Yang-Mills action,
\begin{equation}\label{ym}
S=\int_M Tr(\ast F\wedge F)\,,
\end{equation}
would not give a theory equivalent to General Relativity. 

In the case of the MM theory, the $SO(4,1)$ symmetry is explicitly broken down in to $SO(3,1)$, and the internal Hodge star $\epsilon_{abcd}$ is used (see \cite{wisephd} for a geometrical viewpoint and \cite{RandonoPhD} also Appendix \ref{App-gamma} for the use of Clifford algebras). We therefore have:
\begin{equation}\label{MMYM}
S=-\frac{1}{16\alpha}\int_M Tr\left(i\gamma^5 F\wedge F\right)\,.
\end{equation}
Using (\ref{Fab}) we write the MM action in the following form
\begin{eqnarray}\label{MMYM-comp}
S&=&-\frac{1}{4\alpha}\int_M \epsilon_{abcd}F^{ab}\wedge F^{cd}\\\nonumber
&=&-\frac{1}{2\alpha}\int_M \epsilon_{abcd}\left[\frac{1}{2}R^{ab}\wedge R^{cd}-\frac{1}{\ell^2}e^a\wedge e^b\wedge R^{cd} + \frac{1}{2\ell^4}e^a\wedge e^b\wedge  e^c\wedge e^d\right]\,,
\end{eqnarray}
this action features (in order of appearance) the Euler topological term, the Einstein-Hilbert (EH) term and the cosmological constant term.

The relation between $\Lambda$ and $\ell$ is given by,
\begin{equation}
\frac{\Lambda}{3}=\frac{1}{\ell^2}\,.
\end{equation} 
 
The MM theory can also be formulated as a deformation of a topological theory called $BF$ theory. Topological theories have no \emph{local} degrees of freedom and, they are formulated in a way that does not involve the metric explicitly. In this case, the $BF$ theory for $SO(4,1)$ group is,
\begin{equation}\label{BFMM-top}
S_{}=\int_M \frac{1}{2} B_{IJ}\wedge F^{IJ}\,,
\end{equation}
where $B_{IJ}$ are Lagrange multipliers and $F^{IJ}$ is the curvature of $A^{IJ}$ (\ref{FA}). Note that $BF$ theories (as was mentioned above) do not involve the metric directly in the action. The field equations for the action (\ref{BFMM-top}) are:
\begin{eqnarray}
\nabla_A B_{IJ}=d B_{IJ}+A_I{}^K B_{KJ}+ A_J{}^K B_{IK}&=&0
\\
F^{IJ}&=&0\label{Fzero}\,,
\end{eqnarray}   
so from this equation (\ref{Fzero}) the curvature of $A$ is null. However, a symmetry breaking term may be added making the theory equivalent to GR,
\begin{equation}\label{BFalpha}
S_{}=\int_M\frac{1}{2} B_{IJ}\wedge F^{IJ}-\frac{\alpha}{16} \epsilon_{abcd}B^{ab}\wedge B^{cd}\,,
\end{equation}
or  splitting  $F^{IJ}$ we have,
\begin{equation}\label{BFpart-constr} 
S_{}=\int_M \frac{1}{2}B_{ab}\wedge \left. F^{ab}\right.+ B_a \wedge T^a -\frac{\alpha}{8} \epsilon_{abcd}B^{ab}\wedge B^{cd}\,,
\end{equation}
where we have split $B_{IJ}$ into $B_{ab}$ and $B_{a4}\equiv B_{a}$ in a way similar to the connection $A^{IJ}$ (\ref{MM_A}). 
Variation of (\ref{BFpart-constr}) with respect to $B_{ab}$ gives,
\begin{equation}   
B^{ab}=-\frac{1}{\alpha}\epsilon^{abcd} \left( R_{cd}-\frac{1}{\ell^2}e_c\wedge e_d\right)\,.
\end{equation}
Substituting this in equation (\ref{BFpart-constr}) we have the following action,
\begin{equation}\label{BFpart-action}
S_{} = \int_M -\frac{1}{4\alpha}\epsilon^{abcd}  \left( R_{ab}-\frac{1}{\ell^2}e_a\wedge e_b\right)\wedge \left( R_{cd}-\frac{1}{\ell^2}e_c\wedge e_d\right)+ B_a  \wedge T^a\,,
\end{equation}   
so we have again  an action with the EH term the cosmological term and the Euler topological term. However, unlike the case of action (\ref{MMYM}) in this case the torsion $T^a$ appears explicitly in the action. Variation with respect to $B_a$ gives $T^a=0$. In  \cite{wisephd} this equivalence is studied at the level of the field equations.

The notion of a topological theory is linked with the concept of category. In fact, these theories can be defined as maps between categories that is, as functors. 
A category consists of objects and maps between these objects, called  morphisms. In a category, certain natural laws of composition of morphisms must be satisfied (see \cite{Baez:inv}). In this language, a group is a category with one object and such that all morphism are invertible.  A functor is a map between categories that preserves all the properties of a category.

Categories can be generalized, as higher categories. A (strict\footnote{There is a weaker notion called a bicategory which does not obey all the properties of a strict 2-category. The term \emph{2-category} is sometimes used to mean strict 2-category and this how we will use it.}) 2-category consists of objects, morphisms (maps between the objects) and 2-morphisms (maps between morphisms). Thes1e must satisfy a set of relations between them to form a 2-category see \cite{Baez:inv}. A (strict) 2-group is a (strict) 2-category with one object and such that all morphisms and 2-morphisms are invertible. Two-categories can be further generalized in a similar manner to $n$-categories and $n$-groups.

This way, a possible generalization of the notion of gauge theory consists in writing down theories in which 2-groups  (or $n$-groups in general) play a role similar to gauge groups.  These are called higher gauge theories (HGT). Generalizations of this kind (that is gauge theory for 2-groups) exist, for the YM theory see \cite{Baez:2ym} and for the $BF$ \cite{Martins:2010ry}. Section \ref{sec:hgt} contains a brief summary of these two cases of HGTs .

In Quantum Gravity, a relevant 2-group is the categorical generalization of the Poincaré group $ISO(3,1)$, called the Poincaré 2-group \cite{Baez:inv}. In this work, we construct a categorical generalization of the group $ISO(4,1)$ which we call the \emph{de Sitter 2-group} (Sec. \ref{sec:ds}). Although this 2-group has to our knowledge never been investigated in the context of Quantum Gravity,  the $ISO(4,1)$ group appears in the research literature for example in \cite{PhysRevD.55.2051} and most notably in \cite{Anabalon:2008hi} in this same context of the MM theory.  
  
We generalize the MM theory for this 2-group, first as a higher YM theory (Sec. \ref{sec:2YM}) and subsequently as a higher $BF$ theory (Sec. \ref{sec:2BF}). In both cases, the theory takes the form of an Einstein-Cartan (EC) theory coupled to Kalb-Ramond (KR) fields. A KR filed\footnote{Kalb-Ramond fields are sometimes also called $B$ fields. We will not use this terminology in this work to avoid creating a confusion with the Lagrange multiplier $B$ in the $BF$ action and in the 2-$BF$ action.} is a generalization of the Maxwell (electromagnetic) potential $A_\mu$, it is a two-form $\beta_{\mu\nu}$ and the corresponding curvatures is therefore a three form (we will use the letter $G$ for the curvature of $\beta$). 
Kalb-Ramond fields have appeared in works related to cosmology eg. \cite{universe2010003}, black holes \cite{PhysRevD.101.044014}. They have also bee used in the context of Einstein-Cartan theories \cite{rahaman2005birkhoff} and in modified theories of gravity (see \cite{sym12091573} for a review).

Finally in Sec.\ref{concl} we present our conclusions. 

\section{2-categories 2-groups and higher gauge theories}\label{sec:hgt}
There is an equivalence between (strict) 2-groups and crossed modules (see \cite{Baez:inv} Theorem 2). A crossed module $\left(G,H,\partial , \triangleright \right)$ is made up of two groups $G\,, H$ a homomorphism $\partial: H\rightarrow G $ and a left action of $G$ on $H$ by automorphisms $\triangleright: G \rightarrow Aut(H)$.

Using this crossed module (equivalent to a 2-group) we can construct a gauge theory on a 4-manifold $M$ beginning with a 2-connection $(A\,,\beta)$ where $A$ is a $\mathfrak{h}$ valued 1-form and $\beta$ is a $\mathfrak{g}$ valued 2-form \cite{Baez:inv,Martins:2010ry}.  The forms $A$ and $\beta$ transform under the guage transformations like,
\begin{eqnarray}
A\rightarrow g^{-1}Ag + g^{-1} dg \quad \beta \rightarrow g^{-1} \triangleright \beta
\end{eqnarray}
where $g: M\rightarrow G$. 

Additionally the group $H$ generates the following transformations
\begin{eqnarray}
A\rightarrow A + \partial \eta \quad \beta \rightarrow \beta + d\eta  + A \wedge^\triangleright \eta + \eta \wedge \eta\,,
\end{eqnarray}
where $\eta$ is an $\mathfrak{h}$-valued one-form.

Given this  2-connection, we can construct  a 2-fiber bundle associated to the 2-Lie group $(G, H)$ over the manifold $M$.

In general, the curvature 2-forms for the 2-connection $(A\,,\beta)$ are:
\begin{eqnarray}
\label{calf}\mathcal{F}&=&dA+A\wedge A -\partial \beta\\
\label{calG}\mathcal{G}&=& d\beta + A\wedge ^\triangleright \beta\,.
\end{eqnarray}
These transform under the gauge transformations as,
\begin{eqnarray}
\mathcal{F} \rightarrow g^{-1}\mathcal{F} g \quad \mathcal{G} \rightarrow g^{-1} \triangleright \mathcal{G}
\end{eqnarray}
and 
\begin{eqnarray}
\mathcal{F} \rightarrow \mathcal{F} \quad \mathcal{G} \rightarrow \mathcal{G} + \mathcal{F}\wedge^\triangleright \eta\,.
\end{eqnarray}

In \cite{Baez:2ym} the YM theory is generalized to a higher Yang-Mills theory. This is done by replacing the Lie (gauge) group by a Lie 2-group, we call this theory the 2-Yang-Mills (2YM) theory.

The action for the 2-YM theory can therefore be written
\begin{equation}\label{2ym}
S=\int_M \left\langle \ast \mathcal{F}\wedge \mathcal{F} \right\rangle_\mathfrak{g} +\left\langle	\ast \mathcal{G} \wedge \mathcal{G} \right\rangle_\mathfrak{h}\,.
\end{equation}
Where  $\langle\rangle_{\mathfrak{g}}$ is a $G$-invariant non-degenerate bilinear form in $\mathfrak{g}$ and $\langle\rangle_{\mathfrak{h}}$ is a G-invariant non-degenerate bilinear form in $\mathfrak{h}$.

The wedge product $\wedge$ acts in the following way,
\begin{equation}
    \left\langle\left(a\otimes\mu\right)\wedge \left(b\otimes\nu\right)\right\rangle_\mathfrak{g,h}=\langle a\,,b\rangle_\mathfrak{g,h} \left(\mu\wedge \nu\right)\,,
\end{equation}
where  $a$ and $b$ are elements of one of the Lie algebras $\mathfrak{g,h}$ and $\mu$ and $\nu$ are differential forms.  Note also that $\ast$ is the Hodge operator in  spacetime and acts on the differential form part,
\begin{equation}
\ast \left(a\otimes\mu\right) = \left. a\otimes\ast\mu\right.\,.
\end{equation}

The action for the categorical generalization of the $BF$ theory is,   
\begin{equation}\label{2-BFaction}    
S=\int_M \left\langle B\wedge \mathcal{F} \right\rangle_\mathfrak{g} +\left\langle C \wedge G \right\rangle_\mathfrak{h}\,,
\end{equation}
this is the 2-BF theory\footnote{This theory is sometimes also referred to as the $BFCG$ theory after the fields appearing in the action \cite{MOphd}.}.
The Lagrange multipliers $B$ and $C$ transform as,
\begin{eqnarray}
B \rightarrow g^{-1}B g \quad C \rightarrow g^{-1} \triangleright C\,,
\end{eqnarray}
and 
\begin{eqnarray}
B \rightarrow \left[C\,,\eta \right] \quad C\rightarrow C\,.
\end{eqnarray}

In the next section we construct the \emph{de Sitter 2-group} by categorizing the group $ISO(4,1)$  of inhomogeneous special orthogonal transformations in the de Sitter space.
\section{Construction of the \emph{ de Sitter 2-group}}\label{sec:ds}
In this work, we are interested in the construction of  a 2-group starting from the $ISO(4,1)$ group. It it well known that, for any group of the form $G\ltimes H$ (ie. a semidirect product of $G$ and $H$) a 2-group can be easily constructed (\cite{wisephd} sec. 8.2.2). To build this 2-group,   we will take $G=SO(4,1)$, $H=\mathds{R}^5$, $\partial$ the trivial homomorphism and $\triangleright$ as the representation of $G$ on $H$. The group of morphisms is $\mathds{R}^5$ taken as and abelian group, the group of  2-morphisms will be the semidirect product $SO(4,1)\ltimes \mathds{R}^5$ which is $ISO(4,1)$.
    
We write the connection as $A = A^{IJ}L_{IJ}$ and the two-form $\beta$ as $\beta=\beta^IP_I$

In our case the  Lie Algebra  $\mathfrak{iso}(4,1)$ associated to  $ISO(4,1)$ is written as the semidirect product:
\begin{equation}
\mathfrak{iso}(4,1))=\mathfrak{so}^+(4,1)\ltimes \mathds{R}^5\,,
\end{equation}
of $\mathfrak{so}^+(4,1)$ (the orthochronous special orthogonal Lie algebra of $SO(4,1)$) and the abelian (translation) Lie algebra on $\mathds{R}^5$.

Taking $L_{IJ}=-L_{JI}$ as the generators of $\mathfrak{so}^+(4,1)$ and those of $\mathds{R}^5$ are $P_I$, the Lie bracket of  $\mathfrak{iso}(4,1)$ is,
\begin{eqnarray}\nonumber
\left[L_{IJ}\,,L_{KL}\right]&=&\eta_{LI}L_{JK}- \eta_{JK}L_{IL}+ \eta_{IK}L_{JL} - \eta_{LJ}L_{IK}  \,,  \\
\left[L_{IJ}\,,P_{K}\right]	&=&   \eta_{IK}P_{J}-\eta_{JK}P_{I}\,,	\\
\left[P_I\,,P_J\right]&=& 0\,. \nonumber
\end{eqnarray}

In the particular case of the de Sitter 2-group, $(A\,,\beta)$ is the 2-connection and the associated curvature forms $(F\,, G)$\footnote{In the following, for our particular case, we will drop the calligraphic style of these quantities and write them simply as $F$ and $G$.} are, 
\begin{eqnarray}
	\label{FIJ}F^{IJ}&=&dA^{IJ}+A^I{}_K\wedge A^{KJ}\\
	\label{nbeta}{G}^{I}&=&\nabla_A \beta^I = d\beta^I + A^I{}_J\wedge \beta^J\,.
\end{eqnarray}
The first of these is just the curvature of $A^{IJ}$ which is given by (\ref{FA}). Because of the splitting (\ref{Liesplitt})  the corresponding $SO(3,1)$ and $\mathds{R}^{3,1}$ components are given by (\ref{Fab}) and (\ref{Fa4}) respectively. The second equation (\ref{nbeta}) is the covariant  derivative  of $\beta^I$  and its splitting is given by:
	
\begin{eqnarray}
	\label{nbetaa}\left(\nabla_A \beta\right)^a&=&d\beta^a + \omega^a{}_b\wedge \beta^b + \frac{1}{\ell}e^a{}\wedge \beta_4\,,\\
								 \nonumber &=&\nabla_\omega \beta^a + \frac{1}{\ell}e^a{}\wedge \beta_4\,,\\
	\label{nbeta4}\left(\nabla_A \beta\right)^4&=&d\beta^4 - \frac{1}{\ell}e^b{}\wedge \beta_b\,.
\end{eqnarray}
 
\section{Generalization of MacDowell-Mansouri as a higher Yang-Mills theory}\label{sec:2YM}
In this section we generalize the MM theory using the 2-YM action (\ref{2ym}) for the 2-group we have constructed in the previous section.
 
We note that $\mathfrak{g}\equiv\mathfrak{so(4,1)}$ is equipped with a nondegenerate symmetric bilinear form $\langle\,,\rangle_\mathfrak{g}$ that is invariant under the adjoint action of $G$ and $\mathfrak{h}\equiv\mathds{R}^5$ is likewise equipped with a nondegenerate symmetric bilinear form $\langle\,,\rangle_\mathfrak{h}$ that is invariant under the adjoint action of $H$ and also the action of $G$.
 
Th 2-YM theory for the de Sitter 2-group has the following action,
\begin{equation}
S=\int_M \left.\ast F^{IJ}\wedge  F_{IJ}\right.+ \left.\ast G^I\wedge  G_I  \right.\,.
\end{equation}
This however is neither equivalent to GR not to Einstein-Cartan theory.

In analogy to the process leading from (\ref{ym}) to (\ref{MMYM}) we propose a modification of the above action where the $SO(4,1)$ symmetry is broken in the first term (in practice this is done by replacing $\ast$ by $i\gamma^5$, see Appendix \ref{App-gamma} ) as in the MM case.

We propose the following action, 
\begin{equation}\label{2ymds}
S=\int_M  \left[ -\frac{1}{4\alpha} \epsilon_{abcd}F^{ab}\wedge F^{cd} +\frac{k}{2} \ast G^I\wedge  G_I \right]
\end{equation}
where we have introduced the coupling constants $\alpha$ and $k$ and adjusted constants. Note the similarity of the first term of this equation with (\ref{MMYM-comp}).

Using (\ref{Fab}) we get,
\begin{eqnarray}\label{reeGG}
S=\int_M  -\frac{1}{2\alpha}\left[ \epsilon_{abcd}\left(\frac{1}{2}R^{ab}\wedge R^{cd}-\frac{1}{\ell^2}e^a\wedge e^b\wedge R^{cd}+\right. \right. \nonumber\\
\left. \left. + \frac{1}{2\ell^4}e^a\wedge e^b\wedge  e^c\wedge e^d\right) +\frac{k}{2} \ast G^I\wedge  G_I \right]\,.
\end{eqnarray} 

We now write the differential forms, featuring in  the above action, in components, 
\begin{equation}\label{RGe}
R^{ab}=\frac{1}{2} R^{ab}{}_{\mu\nu}dx^\mu\wedge dx^\nu\,,\quad G^I=\frac{1}{6} G^I{}_{\mu\nu\rho}dx^\mu\wedge dx^\nu\wedge dx^\rho\,,\quad e^a=e^a{}_\mu dx^\mu\,,
\end{equation}
and use this in (\ref{reeGG}) to obtain: 
\begin{eqnarray}\label{MMYMcomp}
S=\int_M \mathrm{d}^4 x\left[ -\frac{1}{16 \alpha}\epsilon_{abcd}\epsilon^{\mu\nu\rho\sigma}\left(R^{ab}{}_{\mu\nu}R^{cd}{}_{\rho\sigma}-\frac{2}{\ell^2}e^a{}_\mu e^b{}_\nu R^{cd}{}_{\rho\sigma}\right.\right.\nonumber\\
\left.\left.+\frac{1}{\ell^4}e^a{}_\mu e^b{}_\nu e^c{}_\rho e^d{}_\sigma\right)-\frac{k |e|}{12\ell^4} G^I{}_{\mu\nu\rho}G_I{}^{\mu\nu\rho}\right]\,,
\end{eqnarray}
where $e=det(e^a{}_\mu)$.

We will also need (\ref{nbetaa}) and (\ref{nbeta4}) in components,
\begin{eqnarray}
G^a{}_{\mu\nu\rho}&=&\nabla_{\!_\omega [\mu}\beta^a{}_{\nu\rho]}+\frac{1}{\ell} e^a{}_{[\mu}\beta^4{}_{\nu\rho]} \label{Ga-comp}\\
G^4{}_{\mu\nu\rho}&=&\partial_{[\mu}\beta^4{}_{\nu\rho]}-\frac{1}{\ell}e^c{}_{[\mu}\beta_c{}_{\nu\rho]}
\end{eqnarray}
where,
\begin{equation}
\nabla_{\!	_\omega \mu}\beta^a{}_{\nu\rho}\equiv \partial_{\mu}\beta^a{}_{\nu\rho} +\omega^a{}_{b\mu}\beta^b{}_{\nu\rho}\,.
\end{equation}  
 
The antisymmetric part of a tensor is defined by,
\begin{equation}
A_{[\mu_1,\mu_2,\ldots,\mu_n]}=\frac{1}{n!}\sum_\pi sgn(\pi)A_{\mu_{\pi(1)},\mu_{\pi(2)},\ldots,\mu_{\pi(n)}]}\,,
\end{equation}
where $\pi$ is a permutation of $\{1,2,\ldots,n\}$ with sign $sgn(\pi)$ and the sum is over all such permutations.

Variation of (\ref{MMYMcomp}) with respect to $e^f{}_\lambda$ gives:
\begin{eqnarray}\label{deltaeym}
&&\frac{1}{4 \alpha}\epsilon_{fbcd}\epsilon^{\lambda\nu\rho\sigma}\left(\frac{1}{\ell} e^b{}_\nu R^{cd}{}_{\rho\sigma}
-\frac{1}{\ell^3} e^b{}_\nu e^c{}_\rho e^d{}_\sigma\right)+\\\nonumber
&&-\frac{k |e|}{12\ell^4}\left( G^I{}_{\mu\nu\rho}G_I{}^{\mu\nu\rho}\ell e_f{}^\lambda - 6G^I{}_{\mu\nu}{}^\lambda G_I{}^{\mu\nu}{}_\gamma\ell e_f{}^\gamma+12G_{[f}{}^{\lambda\nu\rho}\beta_{4]}{}_{\nu\rho}\right)=0\,,
\end{eqnarray} 
where $e_f{}^\lambda$ is the inverse of $e^f{}_\lambda$.

Variation of (\ref{MMYMcomp}) with respect to $\omega^{cd}{}_{\sigma}$
\begin{eqnarray}\label{deltaomega}  
\frac{1}{\ell^2}\epsilon_{abcd}\epsilon^{\mu\nu\rho\sigma}e^a{}_\mu T^b{}_{\nu\rho}=\frac{k\alpha |e|}{4\ell^4}G_{[c}{}
^{\sigma\nu\rho}\beta_{d]\nu\rho}\,,
\end{eqnarray}
where the torsion was defined in (\ref{Fa4}) and may be written in components as,
\begin{equation}    
T^a{}_{\mu\nu}=\frac{2}{\ell}\nabla_{\omega\,[\mu}e^a{}_{\nu]}\,.
\end{equation}
Solving (\ref{deltaomega}) for the torsion gives,
\begin{eqnarray}\label{torsion}
\frac{1}{\ell}T^a{}_{\beta\gamma}=\frac{k\alpha}{6\ell}\left[
2e^a{}_\lambda G_{[\beta}{}^{\lambda\nu\rho} \beta_{\gamma]\nu\rho}
- e^a{}_\beta G_{[\sigma}{}^{\sigma\nu\rho} \beta_{\gamma]\nu\rho}
+e^a{}_\gamma G_{[\sigma}{}^{\sigma\nu\rho} \beta_{\beta]\nu\rho}
\right]\,,
\end{eqnarray}
where we use the shorthand notation,
\begin{equation}
G^a{}_{\mu\nu\rho}=\frac{1}{\ell} e^a{}_\lambda G^\lambda{}_{\mu\nu\rho}\,,\quad
\beta^a{}_{\mu\nu\rho}=\frac{1}{\ell} e^a{}_\lambda \beta^\lambda{}_{\mu\nu\rho}\,.
\end{equation} 
And the variation with respect to $\beta^I{}_{\nu\rho}$ is 
\begin{equation}\label{deltabeym}
\nabla_{\! _A\mu}G_I{}^{\mu\nu\rho}=0\,,
\end{equation}
which may be decomposed in two equations as follows,
\begin{eqnarray}
\nabla_{\omega\mu}G_a{}^{\mu\nu\rho}+e^a{}_\mu G_4{}^{\mu\nu\rho}=0\,,
\\
\partial_{\mu}G_4{}^{\mu\nu\rho}-e^a{}_\mu G_a{}^{\mu\nu\rho}=0\,.
\end{eqnarray}

The Bianchi identity for $A$
\begin{equation}\label{BianF}
\nabla_{A} F^{IJ}=0\,,
\end{equation}
produces the known relations for $R^{ab}$ and $T^a$, namely
\begin{eqnarray}
\nabla_\omega R^{ab}=0\,,\\
\nabla_\omega T^a= R^{a}{}_c\wedge e^c\,.
\end{eqnarray}
And the relation for the derivative of $G^I$,
\begin{equation}\label{BianG}
\nabla_A G^I = F^I{}_J\wedge \beta^J\,,
\end{equation}
reduces to
\begin{equation}
\nabla_\omega \nabla_\omega \beta^a = R^a{}_c\wedge \beta^c\,,
\end{equation}
and $dd\beta^4=0$. Both these relations are identically satisfied.
\section{Generalization of MacDowell-Mansouri as a higher $BF$ theory}\label{sec:2BF}
The generalization of the MM theory may also be carried out using the HGT generalization of the $BF$ theory discussed in Section \ref{sec:hgt}. To do this we first write down a topological theory (based on action (\ref{2-BFaction})) for the 2-group we considered above.

\subsection{Topological theory} \label{subsec:2}
The 2-$BF$ theory for the de Sitter 2-group is given by the following action,
\begin{equation}\label{2BF-action}   
S=\int_M\left( \frac{1}{2} B_{IJ}\wedge F^{IJ}+C_{I}\wedge G^{I}\right)\,,
\end{equation}
where $B_{IJ}$ and $C_{I}$ are Lagrange multipliers, $F^{IJ}$ is given by (\ref{FIJ}) and $G$ by (\ref{nbeta}).
The variables are $\left\{B_{IJ}\,,A^I\,,C_I\,,\beta^I\right\}$, and variation with respect to these gives respectively,
\begin{eqnarray}
F^{IJ}&=&0\,,\nonumber\\
\nabla_A B_{IJ}-2C_{[I}\wedge \beta_{J]}&=&0\,,\\
G^I&=&0\,,\nonumber\\
\nabla_A C_I &=&0\,.\nonumber
\end{eqnarray}

Additionally we have the Bianchi Identities (\ref{BianF})  and (\ref{BianG}) and two additional ones,
\begin{eqnarray}
\label{BianC}\nabla_A \nabla_A C^I &=& F^I{}_J\wedge C^J\,,\\
\nabla_A \nabla_A B^{IJ} &=& F^I{}_K\wedge B^{KJ}+ F^I{}_K\wedge B^{JK}\,.
\end{eqnarray}

To write action (\ref{2BF-action}) in components we use (\ref{RGe}) and,
\begin{eqnarray}\label{Gexpansion}\label{BC}
B^{IJ}=\frac{1}{2}B^{IJ}{}_{\mu\nu}dx^\mu\wedge dx^\nu \,, \qquad C^I=C^I{}_{\mu}dx^\mu \,,
\end{eqnarray}
and we have,
\begin{equation}
S=\int_M d^4x \epsilon^{\mu\nu\rho\sigma}\left(\frac{1}{8}B^{IJ}{}_{\mu\nu}F_{IJ\rho\sigma}+\frac{1}{6}C^I{}_\mu G_{I\nu\rho\sigma}\right)\,.
\end{equation}

Performing a splitting of the spacetime indices $\mu=(0\,,i)\,, i=1,2,3$ we get,
\begin{eqnarray}\label{31action}
S&=&\int_{\mathds{R}\times\Sigma}dtd^3\vec{x}\left(\frac{1}{2}\pi(A)_{IJ}{}^i \dot{A}^{IJ}{}_i
+ \frac{1}{2}\pi(\beta)_{I}{}^{ij} \dot{\beta}^{I}{}_{jk}\right.\nonumber\\
&+&\left.
\frac{1}{2} B^{IJ}{}_{0i} \mathcal{A}_{IJ}{}^i 
+ C^I{}_0 \mathcal{B}_I
+\beta^I{}_{0i}\mathcal{C}_I{}^i
+\frac{1}{2}  A^{IJ}{}_0 \mathcal{D}_{IJ}
\right)\,.
\end{eqnarray}

We see that only (the spatial part of ) $A$ and $\beta$ have time derivatives. Therefore we will consider only these as dynamical variables, (see \cite{Mikovic:2014wfa} for details about this) while $B$ and $C$ will be conjugate momenta. 
These momenta are:
\begin{eqnarray}
\pi(A)_{IJ}{}^k&=&\frac{1}{2}\epsilon^{ijk}B_{IJij}\,,\\
\pi(\beta)_{I}{}^{ij}&=&-\epsilon^{ijk} C_{Ii}\,.
\end{eqnarray}
Here we define $\epsilon^{ijk}\equiv \epsilon^{0ijk}$.

The fundamental Poisson Brackets (PB) brackets between the variables and momenta are defined as:
\begin{eqnarray}
\left\{A^{IJ}{}_i(t,x)\,,\pi(A)_{KL}{}^j(t,x')\right\}&=&2\delta^{I}_{[K}\delta^{J}_{L]}\delta^j_i\delta^{(3)}(x-x')\,,\\
\left\{\beta^I{}_{ij}(t,x)\,,\pi(\beta)_{J}{}^{kl}(t,x')\right\}&=&2\delta^I_J\delta_{[i}^{k}\delta_{j]}^{l}\delta^{(3)}(x-x')\,.
\end{eqnarray}

From (\ref{31action}) we can read off the following  constraints:
\begin{eqnarray}
\label{A}\mathcal{A}_{IJ}{}^i&=&\frac{1}{2}\epsilon^{ijk}F_{IJ jk}\,,\\
\mathcal{B}_I&=&\frac{1}{6}\epsilon^{ijk}G_{Iijk}\,,\\
\label{C}\mathcal{C}_I{}^i&=&\nabla_{Aj}\pi(\beta)_I{}^{ji}\,,\\
\mathcal{D}_{IJ}&=&\nabla_{Ak}\pi(A)_{IJ}{}^k-\frac{1}{2}\epsilon^{ijk}\left(C_{Ii}\beta_{Jjk}-C_{Ji}\beta_{Ijk}\right)\,.
\end{eqnarray}
Not all the components of these are independent. Using (\ref{BianF}) we find,
\begin{equation}\label{nablaA}
\nabla_{Ai}\mathcal{A}_{IJ}{}^i=0\,,
\end{equation}
thus reducing the number of independent components of (\ref{A}). Furthermore, using (\ref{BianC}) we find (taking the spacial part) that,
\begin{equation}\label{nablaC}
\nabla_{A i}\mathcal{C}_I{}^i = \frac{1}{2} \epsilon^{ijk}F_{IJ ij} e^J{}_k\,,
\end{equation}
this means that the components of (\ref{C}) are not all independent.

These constraints are First Class as can  be seen from their PBs,
\begin{eqnarray}  
\left\{ \mathcal{B}_I (t,x) \,,  \mathcal{C}_J{}^i(t,x') \right\}&=& - \mathcal{A}_{IJ}{}^i (t,x){} \delta^{(3)}(x-x')\,,\nonumber\\
\left\{ \mathcal{B}^I (t,x) \,, \mathcal{D}_{JK} (t,x') \right\}&=& 2 \delta^I_{[J}\mathcal{B}_{K]} (t,x)  \delta^{(3)}(x-x')\,,\nonumber\\ 
\left\{ \mathcal{A}^{IJi} (t,x) \,, \mathcal{D}_{KL} (t,x') \right\}&=&-4  \delta^{[I}_{[K}\mathcal{A}^{J]}{}_{L]}{}^{i}(t,x)\,, \delta^{(3)}(x-x')\,,\\
\left\{ \mathcal{D}_{IJ} (t,x) \,, \mathcal{D}^{KL} (t,x') \right\}&=&-4 \delta^{[K}_{[I}\mathcal{D}^{L]}{}_{J]}(t,x)\delta^{(3)}(x-x')\,,\nonumber\\
\left\{ \mathcal{D}_{IJ} (t,x) \,,  \mathcal{B}^K(t,x') \right\}&=& -2 \delta^K_{[I}\mathcal{B}_{J]}(t,x)   \delta^{(3)}(x-x')\,.\nonumber
\end{eqnarray}

The variables have the following independent components:
\begin{center}

\begin{tabular}{|c|c|c|c|}
\hline 
 field & $A^{IJ}{}_i$ & $\beta^I{}_{ij}$& Total \\ 
\hline 
number of components & 30  & 15 & 45\\ 
\hline 
\end{tabular} 
\end{center}
And the number of independent  components of the First Class constraints is,
\begin{center}
\begin{tabular}{|c|c|c|c|c|c|}
\hline 
constraint & $\displaystyle \mathcal{A}_{IJ}{}^i$ & $\mathcal{B}_I$ & $\mathcal{C}_I^i$ & $\mathcal{D}_{IJ}$ & Total\\ 
\hline 
 number of components & 30-10 & 5 & 15 - 5 & 10 & 45 \\ 
\hline 
\end{tabular} 
\end{center}
where we have subtracted the number of components corresponding to (\ref{nablaA}) and (\ref{nablaC}).

Using the relation \cite{MOphd} between the number of local degrees of freedom $n$, the number of local independent field components $N$, the number of local independent FC constraint components $F$ and the number of local Second Class constraint components $S$,
\begin{equation}
n=N-F-\frac{S}{2}\,,
\end{equation} 
and since there are no Second Class constraints $S=0$,  we conclude that there are no local degrees of freedom. 

There is a relation between action (\ref{2BF-action}) and the $BF$ theory for the  $ISO(4,1)$ group. This relation was studied in \cite{Mikovic:2014wfa} for the $ISO(3,1)$ case. 
Using an integration by parts the action (\ref{2BF-action}) can be written
\begin{equation}\label{}
S=\int_M \frac{1}{2} B_{IJ}\wedge F^{IJ} + \beta_{I} \wedge  \nabla_A C^{I}-d\left(C_I\wedge\beta^I\right)\,,
\end{equation}
and discarding the total derivative $d\left(C_I\wedge\beta^I\right)$ we have,
\begin{equation}\label{}
S=\int \frac{1}{2} B_{MN}\wedge F^{MN} \,,
\end{equation}
where $M,N=0	\,,\ldots\,,5\,.$\footnote{We make and exception here to the conventions we use in the rest of this work.} we have used the following identifications:
\begin{eqnarray}
 B_{I5}&=&\beta_{I}\nonumber \\
 A_{I5}&=&C_{I}\\
F^{I5}&=& dC^{I} + A^{I}{}_{K}\wedge C^K\nonumber
\end{eqnarray}
This establishes the fact that theory given by (\ref{2BF-action}) can be written as a $BF$ theory for the $ISO(4,1)$ group. 
\subsection{Constrained theory}   
The action for the constrained theory may be written in the form:
\begin{equation}\label{s2mm}
S_{2MM}=S_{BF}\left[B\,,\omega\,,e\right]+S_{CG}\left[C\,,\beta\,,\omega\,,e\,, \phi\right]\,.
\end{equation}
For the first part of this action $S_{BF}$, we modify the symmetry breaking term in (\ref{BFalpha}). Instead of having just $\epsilon_{abcd}B^{ab}\wedge B^{cd}$, we now have:
\begin{equation}
B^{IJ}\wedge \tilde B_{IJ}\,,
\end{equation}
\begin{eqnarray}
\tilde B_{ab}&=&\epsilon_{abcd} B^{cd}+2B_{ab}\,,\\
\tilde B_{a4}&=& 4B_{a4}\,.
\end{eqnarray}
Therefore the action $S_{BF}$ is
\begin{eqnarray}\label{BF-action}
S_{BF}&=&\int_M\left[\frac{1}{2} B^{IJ} \wedge F_{IJ}-\frac{\alpha}{32} B^{IJ}\wedge \tilde B^{IJ}\right]\,.
\end{eqnarray}  
Splitting the Lie algebra indices we have:
\begin{eqnarray}\label{BF-action-split}
S_{BF}&=&\int_M\left[\frac{1}{2} B^{ab} \wedge F_{ab}-\frac{\alpha}{32}\left(\epsilon_{abcd}B^{ab}\wedge B^{cd}+2B^{ab}\wedge B_{ab}\right)\right.\nonumber\\
&+& \left.\frac{1}{\ell} B^{a4}\wedge T_a -\frac{\alpha}{4}B^{a4}\wedge B_{a4}\right]\,.
\end{eqnarray}   

Regarding the second part of (\ref{s2mm}), the mechanism is different. Instead of adding a symmetry breaking term,  we use an auxiliary field $\phi_{Ib}$ that acts as a Lagrange multiplier. Variation with respect to this field will enforce a constraint in the $C$ field as we will see shortly.  This part of the action reads:
\begin{equation}\label{CG-action} 
S_{CG}=\int_M \left[C^I\wedge G_I - \phi_{Ib}\left(\frac{k}{\ell} G^I \wedge e^b - C^I\wedge \frac{1}{3!\ell^3}\epsilon^b{}_{cde}e^c\wedge e^d \wedge e^e\right)\right]\,.
\end{equation}

The variation of $S_{2MM}$ with respect to $B^{ab}$ reduces to the variation of $S_{BF}$. This gives, 
\begin{equation}\label{delta_Bab}
F_{ab}=\frac{\alpha}{8}\left(\epsilon_{ab}{}^{cd}B_{cd}+2B_{ab}\right)\,,
\end{equation}
solving for $B_{cd}$ we find,
\begin{equation}\label{Bab}
B_{cd}=-\frac{1}{\alpha}\left(\epsilon^{ab}{}_{cd}F_{ab}-2F_{cd}\right)\,.
\end{equation}
Furthermore, variation of (\ref{BF-action}) with respect to $B^{a}$ gives,
\begin{equation}\label{Ba}
B_a=\frac{2}{\alpha\ell}T_a\,.
\end{equation}
Using (\ref{Bab}) and (\ref{Ba}) equation (\ref{BF-action}) becomes,
\begin{equation} 
S_{BF}=\int -\frac{1}{4\alpha}\left( \epsilon_{abcd}F^{ab}\wedge F^{cd}-2F^{ab}\wedge F_{ab}\right)+\frac{1}{\alpha  \ell^2}T^a\wedge T_a\,,
\end{equation}
and with the help of  (\ref{Fab}) we have
\begin{eqnarray}
 S_{BF}=\int_M &-&\frac{1}{4\alpha} \left(\epsilon_{abcd}R^{ab}\wedge R^{cd} -\frac{2}{\ell^2}\epsilon_{abcd}R^{ab}\wedge e^c \wedge e^d+
\right. \nonumber \\ 
&+&\left. \frac{1}{\ell^4}\epsilon_{abcd}e^a \wedge e^b \wedge e^c \wedge e^d \right) +\\
&+& \frac{1}{2\alpha}R^{ab}\wedge R_{ab}-\frac{1}{\alpha \ell^2}\left( R_{ab}\wedge e^a\wedge e^b - T^a\wedge T_a \right)\nonumber\,.
\end{eqnarray}
 
This action features the following $SO(3,1)$ topological terms:
\begin{eqnarray}
R^{ab}\wedge R_{ab} &\qquad & \mbox{Pontrjagin term}\nonumber \\
R_{ab}\wedge e^a\wedge e^b - T^a\wedge T_a &\qquad&  \mbox{Nieh-Yan  term}\nonumber \\
\epsilon_{abcd}R^{ab}\wedge R^{cd} &\qquad& \mbox{Euler term}
\end{eqnarray}
it features also, 
\begin{eqnarray}
\epsilon_{abcd}R^{ab}\wedge e^c \wedge e^d &\qquad & \mbox{Einstein-Palatini term}\nonumber\\
\epsilon_{abcd}e^a \wedge e^b \wedge e^c \wedge e^d & \qquad & \mbox{cosmological constant term}\,.
\end{eqnarray}
Note that, for theories with non-zero torsion (which is our case) the Holst term $R_{ab}\wedge e^a\wedge e^b$ is not topological. In this case the topological invariant is $R_{ab}\wedge e^a\wedge e^b - T^a\wedge T_a $ the Nieh-Yan  term.

As for the second part of the action (\ref{CG-action}),  variation with respect to $\phi_{Ib}$ gives,
\begin{equation}\label{cons-eq}
\frac{k}{\ell} G^I \wedge e^b - C^I\wedge \frac{1}{3!\ell^3}\epsilon^b{}_{cde}e^c\wedge e^d \wedge e^e=0\,,
\end{equation}
the solution of this equation is (see Appendix \ref{sol-cons-eq} for a proof using the inverse of the co-tetrad $e^a{}_\mu$)

\begin{equation}\label{Ceq}
C^I=k\ast G^I\,.
\end{equation}

This can also be found writing the forms in the $e^a$ basis
\begin{equation}
G^I=\frac{1}{3!\ell^3}G^I_{bcd}e^b\wedge e^c \wedge e^d\,,\quad C^I=\frac{1}{\ell} C^I{}_{b}e^b\,.
\end{equation} 
equation (\ref{cons-eq}) becomes,
\begin{eqnarray}
\frac{k}{3!\ell^4} G^I{}_{cde}e^c\wedge e^d\wedge e^e \wedge e^b-\frac{1}{3!\ell^4} C^I{}_f e^f\wedge \epsilon^b{}_{cde}e^c\wedge e^d \wedge e^e&=&0\nonumber\\
\frac{|e|}{\ell^4}\left(-\frac{k}{3!}\epsilon_b{}^{cde}G^I{}_{cde}-C^I{}_f \frac{1}{3!} \epsilon^{fcde}\epsilon{}_{bcde}\right)&=&0\nonumber\\
-\frac{k}{3!} \epsilon_b{}^{cde}G^I{}_{cde}+C^I{}_f \delta^f_b&=&0\,,
\end{eqnarray}
and finally we get
\begin{equation}
C^I{}_b=\frac{k}{3!}\epsilon_b{}^{cde}G^I{}_{cde}\,.
\end{equation}
Using this in the action (\ref{CG-action}) we get,

\begin{equation}\label{CG-action-simp} 
S_{CG}=\int_M k\left.\ast G^I\wedge G_I \right.  
\end{equation}
  
Variation of the action (\ref{s2mm}) gives,
\begin{eqnarray}
\delta_\omega:&&\quad -\nabla_\omega B_{ab}- \frac{1}{\ell} B_{[a}\wedge e_{b]} +C_{[a}\wedge \beta_{b]}+\frac{k}{\ell}\phi_{[a|c|} e^c\wedge \beta_{b]}=0\label{deltaomegabf}\\
\delta_e:&& -\frac{2}{\ell} B_{ab}\wedge e^a -\nabla_\omega B_b +2 C_{[b}\wedge \beta{}_{4]}\label{delta_e}\nonumber\\ 
&& -k\phi_{Ib}G^I 
+\frac{2k}{\ell}\phi_{[b |a|} \wedge e^a\wedge \beta_{4]} +\frac{1}{2\ell^2}\phi_{Ia}C^I\wedge \epsilon^a{}_{cdb}e^c\wedge e^d=0\\
\delta_{C}: &&\quad G_I + \phi_{Ib} \frac{1}{3!\ell^3}\epsilon^b{}_{cde}e^c\wedge e^d \wedge e^e=0\label{delta_c}\\
\delta_\beta:&& \quad \nabla_A \left( C^I + k\varphi^I\right)=0\label{deltabeta}\,,
\end{eqnarray}
where $\varphi^I =\phi^I{}_b e^b$.

Additionally we have, equations (\ref{delta_Bab})   (\ref{Ba})   (\ref{cons-eq}) resulting from the variation with respect to $B^{ab}$, $B^{a}$ and $\phi_{Ib}$.

Equation (\ref{delta_c}) can be solved (see Appendix \ref{sed:solfi}). The solution is,
\begin{equation}\label{varphi}
\varphi^I = \ast G^I\,.
\end{equation}
Using this equation and (\ref{Ceq}) equation (\ref{deltabeta}) becomes
\begin{equation} 
\nabla_A \ast G^I=0
\end{equation}
this is equation (\ref{deltabeym}) we obtained in the higher Yang-Mills formulation of the MM theory.
 
With the help of  (\ref{Bab}) (\ref{Ba}) (\ref{Ceq})  and (\ref{varphi}) equation (\ref{deltaomegabf}) becomes
\begin{equation}
\frac{1}{\ell^2}\epsilon_{abcd}e^c \wedge T^d - {2\alpha k} \ast G^I_{[a} \wedge \beta_{I b]}=0
\end{equation}
this is equivalent to equation (\ref{deltaomega}).

Equation (\ref{delta_e}) can be written,
\begin{equation}
\frac{1}{\alpha}\left(\frac{1}{\ell}\epsilon_{abcd}R^{ab}\wedge e^c- \frac{1}{\ell^3}e^a\wedge e^b \wedge e^c\right)+4\ast G_{[b}\wedge \beta_{4]} - \ast G^I{}_b \wedge  G_I -3 G_I{}_b\wedge \ast G^I =0\,.
\end{equation}
And using the identity for the interior product $\iota$ of the wedge product of the  $p$-form $\beta$ and the $q$-form $\gamma$, with the vector $X$,
\begin{equation}
\iota_X \left(\beta \wedge \gamma\right)=\iota_X \left(\beta\right) \wedge \gamma+ \left(-1\right)^p\beta \wedge \iota_X \left(\gamma\right)
\end{equation}

in the term $- \ast G^I{}_b \wedge  G_I$ we have,
\begin{equation}
\frac{1}{\alpha}\left(\frac{1}{\ell}\epsilon_{abcd}R^{ab}\wedge e^c- \frac{1}{\ell^3}e^a\wedge e^b \wedge e^c\right)+4\ast G_{[b}\wedge \beta_{4]} +( G^I \wedge  \ast G_I){}_b-4 G_I{}_b \wedge \ast G^I =0
\end{equation}
where we have,
\begin{equation}
\ast G^I = (\ast G^I){}_b  e^b\quad  G^I = G^I {}_b  e^b \quad \ast G^I \wedge G_I=(\ast G^I \wedge G_I){}_b e^b
\end{equation}
This is equivalent to equation  (\ref{deltaeym}) of the higher YM formulation. 
\section{Conclusions}\label{concl}
In this work, we study the generalizations of the MM theory in the context of strict 2-groups. The MM theory, relies on the use of a symmetry group containing the Lorentz group. In our case we use  $SO(4,1)$. The Lie algebra associated to this group is then decomposed as the direct sum of two vector spaces namely $\mathfrak{so}(4,1)\cong \mathfrak{so}(3,1) \oplus \mathds{R}^{3,1}$. The MM theory may be cast in a form similar to the  Yang-Mills theory. In this case, the $SO(4,1)$ symmetry is explicitly broken by the use of $\epsilon_{abcd}$ (\ref{MMYM-comp}). The MM theory may also be formulated  as a $BF$ theory to which a symmetry breaking term (once again involving $\epsilon_{abcd}$) is added (\ref{BFalpha}). 

The construction of a higher gauge theory requires a strict 2-group. We use the categorification of the $ISO(4,1)$ group (Section \ref{sec:ds}) which we call the \emph{de Sitter 2-group}. 

To generalize the MM theory as a Yang-Mills-like theory, we use the 2-YM theory proposed in \cite{Baez:2ym}. In order to recover a theory of gravity close to GR, we explicitly break the $SO(4,1)$ symmetry using $\epsilon_{abcd}$. The result is an EC theory of gravity coupled to Kalb-Ramond fields. The torsion in this theory is related to these KR fields by (\ref{torsion}).

 As for the $BF$ formulation of the MM theory, its generalization makes use of the 2-$BF$ theory (\ref{2-BFaction}). This theory, when specialized to the \emph{de Sitter 2-group} produces a topological theory. We establish this by using a spacetime decomposition of the action and studying the number of First Class constraints (Subsection \ref{subsec:2}). We then proceed to add two constraint terms to this topological theory, therefore constructing an Einstein-Cartan theory  given in equations (\ref{s2mm}) (\ref{BF-action}) and (\ref{CG-action}). The theory in the $BF$ form is equivalent to the one in the YM form except for the extra Nieh-Yan topological term. We note that, both forms of the generalization include the cosmological constant term. 

The canonical quantization of these theories, requires their Hamiltonian analysis to be carried out. This is work in progress and will appear elsewhere.

\appendix
\section{Identities involving Gamma matrices}\label{App-gamma}
We take the Dirac gamma matrices as obeying,
\begin{equation}
\left\{\gamma^a, \gamma^b\right\}=2\eta^{ab}\,,
\end{equation}
and we define $\gamma_5$ as,
\begin{equation}
\gamma_5 = i\gamma^0\gamma^1\gamma^2\gamma^3=\frac{i}{4!}\epsilon_{abcd}\gamma^a\gamma^b\gamma^c\gamma^d\,.
\end{equation}
with the convention $\epsilon_{0123}=+1$. 

It is then easy to prove,  
\begin{equation}\label{igamma}
Tr\left(i\gamma_5\gamma_a\gamma_b\gamma_c\gamma_d\right)=4\epsilon_{abcd}\,.
\end{equation}

This allows us to use $i\gamma_5$ as a sort of `Hodge star operator'. We have the following MM action:
\begin{eqnarray}
S&=&-\frac{1}{16\alpha}\int_M Tr\left(i\gamma_5F^{IJ}\wedge F^{KL} \gamma_{I}\gamma_{J}\gamma_{K}\gamma_{L}\right)\\
&=&-\frac{1}{4\alpha}\int_M \epsilon_{abcd}F^{ab}\wedge F^{cd}\,.
\end{eqnarray}
Note that (\ref{igamma}) is the only non-zero combination, given the trace properties of gamma matrices. This is eq. (\ref{MMYM-comp}) in the main text. 

The $BF$ action can be written,
\begin{eqnarray}
S&=&-\frac{1}{16}\int Tr\left(B^{IJ}\wedge F^{KL}\gamma_I\gamma_J\gamma_K\gamma_L\right)\\
  &=&-\frac{1}{16}\int Tr\left(B^{ab}\wedge F^{cd}\gamma_a\gamma_b\gamma_c\gamma_d\right)-\frac{4}{16}\int Tr\left(B^{4a}\wedge F^{4b}\gamma_5\gamma_a\gamma_5\gamma_b\right)\nonumber
\end{eqnarray}
where $\gamma_{I=4}\equiv \gamma_5$. Using the fact that $(\gamma_5)^2=I$ and that it anticommutes  with the other $\gamma_a$, together with the trace identities for two and four gamma matrices,
\begin{eqnarray}
Tr\left(\gamma_a\gamma_b\right)&=&4\eta_{ab}\\
Tr\left(\gamma_a\gamma_b\gamma_c\gamma_d\right)&=&4\left(\eta_{cd}\eta_{ab}-\eta_{bd}\eta_{ac}+\eta_{ad}\eta_{bc}\right)
\end{eqnarray}
we have,
\begin{eqnarray}  
S&=&\int\left(\frac{1}{2} B^{ab}\wedge F_{ab}+B^{4a}\wedge F_{4a}\right)\nonumber\\
&=&\int\frac{1}{2} B^{IJ}\wedge F_{IJ}\,.
\end{eqnarray}
these are the first two terms in (\ref{BFpart-constr}).

The third term can be written as,
\begin{eqnarray}
S'&=&-\frac{\alpha}{4^3}\int tr\left(B^{IJ}\wedge B^{KL}i\gamma_5\gamma_I\gamma_J\gamma_k\gamma_L\right)\\
  &=&-\frac{\alpha}{16}\int \epsilon_{abcd}\left.B^{ab}\wedge B^{cd}\right.\,.
\end{eqnarray}

So the full action is, 
\begin{eqnarray}
S=-\int Tr\left(\frac{1}{4^2}B^{IJ}\wedge F^{KL}\gamma_I\gamma_J\gamma_K\gamma_L-\frac{\alpha}{4^3}B^{IJ}\wedge B^{KL}i\gamma_5\gamma_I\gamma_J\gamma_k\gamma_L\right)\,.
\end{eqnarray}

The action for the generalization of the MM as a higher YM theory (\ref{2ymds}) may be written,
\begin{eqnarray}
S_{MM}=-\int_M Tr\left(\frac{1}{16\alpha}i\gamma_5F^{IJ}\wedge F^{KL} \gamma_{I}\gamma_{J}\gamma_{K}\gamma_{L}+ k G^I\wedge\ast G^J \gamma_I\gamma_J\right)\,.
\end{eqnarray}

\section{Solution of the `constraint equation'}\label{sol-cons-eq}
In this section we solve the `constraint equation' (\ref{cons-eq}) using the component notation. 

Using (\ref{RGe}) and (\ref{BC}), equation (\ref{cons-eq}) becomes,
\begin{equation}\label{cons-eq-comp}
\frac{k}{6\ell}\epsilon^{\mu\nu\rho\sigma}G^I{}_{\mu\nu\rho}e^b{}_{\sigma}- \frac{1}{6\ell^3}\epsilon^{\mu\nu\rho\sigma}C^I{}_{\mu} \epsilon^b{}_{cde}e^c{}_\nu e^d{}_\rho  e^e{}_\sigma=0\,,
\end{equation}
contracting with $e_{b\lambda}$ (note that by doing this we are implicitly using the metric $g_{\sigma \lambda} = \frac{1}{\ell^2}e^b{}_\sigma e_{b\lambda}$) we get,

\begin{eqnarray}
\frac{k}{6}\epsilon^{\mu\nu\rho}{}_\lambda G^I{}_{\mu\nu\rho}- C^I{}_{\mu} \frac{1}{6\ell^4}\epsilon^{\mu\nu\rho\sigma}\left( \epsilon^b{}_{cde}e_{b\lambda}e^c{}_\nu e^d{}_\rho  e^e{}_\sigma\right)&=&0\nonumber\\
\frac{k}{6}\epsilon^{\mu\nu\rho}{}_\lambda G^I{}_{\mu\nu\rho}- C^I{}_{\mu}\frac{|e|}{\ell^4} \frac{1}{6}\epsilon^{\mu\nu\rho\sigma}\epsilon_{\lambda\nu\rho\sigma}&=&0\nonumber\\
\frac{k}{6}\epsilon^{\mu\nu\rho}{}_\lambda G^I{}_{\mu\nu\rho}+ C^I{}_{\lambda}\frac{|e|}{\ell^4} &=&0\nonumber
\end{eqnarray}
and finally we get
\begin{equation}
C^I{}_{\lambda}=\frac{k\ell^4}{6|e|}\epsilon_\lambda{}^{\mu\nu\rho}{} G^I{}_{\mu\nu\rho}
\end{equation}
\section{Solution of equation (\ref{delta_c})}\label{sed:solfi}
The field equation (\ref{delta_c}) can be written in component form, using (\ref{Gexpansion}). It reads,
\begin{eqnarray}\label{delta_c_comp}
\frac{1}{3!}\epsilon^{\alpha\mu\nu\rho}G^I{}_{\mu\nu\rho}+\phi^{Ib}\frac{1}{3!\ell^3}\epsilon^{\alpha\mu\nu\rho}\epsilon_{bcde}e^c{}_\mu e^d{}_\nu e^e{}_\rho=0\,,
\end{eqnarray}
Using $e_{b'}{}^\beta$ we can write,
\begin{equation}
\phi^{Ib}=\phi^{Ib'}\ell e_{b'}{}^\beta \frac{1}{\ell} e^{b}{}_\beta\equiv \varphi^{I\beta} \frac{1}{\ell} e^{b}{}_\beta
\end{equation}
with this equation (\ref{delta_c_comp}) becomes
\begin{eqnarray}\label{}
\frac{1}{3!}\epsilon^{\alpha\mu\nu\rho}G^I{}_{\mu\nu\rho}+\varphi^{I\beta}\frac{1}{3!\ell^4}\epsilon^{\alpha\mu\nu\rho}\epsilon_{bcde}  e^{b}{}_\beta e^c{}_\mu e^d{}_\nu e^e{}_\rho&=&0\nonumber \\
\frac{1}{3!}\epsilon^{\alpha\mu\nu\rho}G^I{}_{\mu\nu\rho}-\varphi^{I\beta}\delta_\beta^\alpha\frac{|e|}{\ell^4}&=&0\nonumber
\end{eqnarray}
and finally  we get, 
\begin{equation}
\varphi^{I}{}_\beta = \frac{\ell^4}{6|e|} \epsilon_\beta{}^{\mu\nu\rho}G^I{}_{\mu\nu\rho}
\end{equation}
or equivalently 
\begin{equation}
\phi^{Ia}{}_b e^b{}_\beta  = C^I{}_\beta
\end{equation}


\section*{Acknowledgments}
This work was supported by the Fundação para a Ciência e a Tecnologia (FCT), under the contract with the reference CEECIND/03053/2017.
We thank also A. Mikovi\'c for stimulating conversations on this subject, J. Mourão for a review of the manuscript and John Huerta for discussion on the various aspects of this work.

\bibliographystyle{unsrt}
\bibliography{refs}   

\end{document}